\documentclass[mathleft]{an}
\usepackage{graphicx}
\usepackage{times}
\begin{document}
\Pagespan{789}{}
\Yearpublication{2006}%
\Yearsubmission{2006}%
\Month{11}%
\Volume{999}%
\Issue{88}%

\title{Magnetic properties of three cool CP stars with strong fields\thanks{Work is based on the analysis of data collected with the Russian 6-m telescope.}}

\author{E.A. Semenko\thanks{Corresponding author:
  {sea@sao.ru}\newline}
\and  L.A. Kichigina
\and E.Yu. Kuchaeva
}
\titlerunning{Magnetic properties of three cool CP stars}
\authorrunning{E.A. Semenko, L.A. Kichigina \& E.Yu. Kuchaeva}
\institute{
Special Astrophysical Observatory of the Russian Academy of Sciences, Nizhnij Arkhyz, Russia, 369167}

\received{}
\accepted{}
\publonline{later}


\abstract{%
The study of magnetic fields of  cool chemically peculiar stars with effective temperatures less than 
10\,000\,K is very important to understand  the nature of their magnetism. We present new results of a long-term spectroscopic monitoring of the well-known magnetic star HD\,178892. The analysis of spectra taken with the Russian 6-m telescope has revealed a periodic variation of the surface magnetic field from 17 to 23\,kG. A revised rotational period of HD 178892 was extracted from the mean longitudinal field:  8.2549 days. We have continued the study of the components of the magnetic binary  BD $+40^{\circ}175$  started by V.\,Elkin at SAO RAS. Our measurements of magnetically splitted lines in the spectra of each component show the presence of strong magnetic fields in both components. 
The surface field in the case of the component A was about 14\,kG at three different epochs. The component B possesses a slightly weaker field: $B_\mathrm{s}$ varies from 9 to 11\,kG. A preliminary analysis of the chemical abundances allows us to make an assumption about the roAp nature of both components of BD $+40^{\circ}175$.}

\maketitle

\section{Introduction}
An interest in  searching and studying stars possessing  very strong magnetic
fields was always existing. Now, after more than 60 years since the discovery of the
magnetic field of 78 Vir by Babcock, we know about more than two dozens of stars
with surface magnetic fields exceeding 10\,kG. Among these objects, HD 215441, also
known as `Babcock's star', is the leader in magnetic field strength~(Babcock
1960). Recently, Elkin et al.~(2010a,b) found a rival for Babcock's star~-- HD
75049, the magnetic field of which reaches 30 kG on the surface. Another strongly
magnetic stars are HD 37776 (Borra \& Landstreet 1979; Kochukhov et al. 2011), HD 137509 (Kochukhov 2006), and some other objects. However, these stars are mostly hot (${T_\mathrm{eff}>10\,000}$ K) and have silicon and helium abundance anomalies.

As for strongly magnetic stars with effective temperatures below 10\,000\,K, the first detection of surface magnetic field greater than 20 kG was done by Babel \& North (1997) for the main component of the binary system BD\,$+40^{\circ}175$. Elkin~(1999) established the magnetic nature of the second component of BD\,$+40^{\circ}175$, that like  the main component is characterized by a low temperature and enhanced lines of Sr, Sm, Eu, and Gd. 

Circularly polarized spectra of BD\,$+40^{\circ}175$AB showed that both stars 
 possess a magnetic field that varies from about $-$2 to $-$3.5\,kG and from 0.7 to 2.6\,kG for the A
 and B component, respectively. This means that the sign of the magnetic field $B_\mathrm{e}$  is
 opposite for component B. Unfortunately, all results presented by Elkin are completely based on
 medium-resolution long-slit spectra that makes any detailed study of chemical abundances
 and magnetic field geometry for both components of BD\,$+40^{\circ}175$  impossible.

The next two cool magnetic stars with strong surface magnetic field were found approximately at the same time. Observing with the 6-m telescope BTA, Elkin et al.~(2003) have measured in spectra of 
HD 178892  magnetically induced shifts between the left and right circularly polarized lines which correspond to $B_\mathrm{e}$ of more than 6--8\,kG. This means that the surface magnetic field of 
HD 178892 should be exceptionally high. Later, this conclusion was confirmed in the paper by Ryabchikova et al.~(2006). From the measurements of magnetically split lines they established the presence of a strong magnetic field ($\langle B_\mathrm{s}\rangle =17.5$\,kG).
From photometrical data a rotational period ${P=8.\!\!^\mathrm{d}2478}$ was found. Using the effective temperature of the star ($T_\mathrm{eff}=7700$\,K), extracted from Geneva photometry and hydrogen lines, Ryabchikova et al. performed an abundance analysis of the stellar atmosphere. They found an abundance pattern typical for cool rapidly pulsating (ro)Ap stars. Assuming a dipolar structure for the  magnetic field, the authors have determined the dipole parameters. According to their results, the polar strength of the dipole should be about 23\,kG, and the value of $\langle B_\mathrm{s}\rangle$ should lie between 15.1 and 21.1\,kG.

The magnetic field of the Ap star HD 154708 was discovered by Hubrig et al.~(2005). They measured spectra collected with FORS1 in polarimetric mode. From  the hydrogen lines the authors determined a mean longitudinal field of 7.5\,kG. Measured from UVES spectra, the mean surface magnetic field ($\langle B_\mathrm{s}\rangle$) made HD 154708 the second strongest magnetic star after  Babcock's star and the first~-- amongst the cool Ap stars. Hubrig et al. have noticed a relatively low temperature~($T_\mathrm{eff}=6800$\,K) and, hence, the position of HD 154708 in the HR diagram makes
 it a good candidate for a roAp star. Later, rapid pulsations with a period of 8 minutes were found by Kurtz et al.~(2006). An unusually low amplitude of pulsation of rare-earth elements or REEs~($\sim$ 60\,m\,s$^{-1}$), measured from a series of high-resolution \'echelle spectra, explained the authors as a result of the influence of an extremely strong magnetic field in  the atmospheric layers. 

The poorly known work by Nesvacil et al.~(2008) was aimed on a detailed study of the individual chemical abundances in the atmosphere of HD 154708. It was shown by the authors that heavy and rare-earth elements are strongly  overabundant. Also, the presence of the 
so-called Pr-Nd anomaly is clearly seen. The latest paper by Hubrig et al.~(2009), related to the study of HD 154708, was aimed on determining the geometry of the stellar magnetic field. In particular, in this work a new, more accurate rotational period was established and parameters of the dipole, describing observable magnetic quantities, were found. Later, we shall compare our results 
for HD 178892 with the earlier published ones.

The last cool Ap star with a strong magnetic field was discovered recently by Elkin et al.~(2009): For  BD $+00^{\circ}4535$ they measured an average surface modulus of the magnetic field that happened to be very high. The value of ${\langle B_\mathrm{s}\rangle=21}$\,kG has put this star to the set of five cool Ap stars with the strongest magnetic fields. The authors have performed a detailed study of magnetic and chemical properties of the star, but all their conclusions are based on just two spectra obtained with FEROS, installed at the 2.2-m telescope of ESO's La Silla Observatory. Neither the spectral nor magnetic variability of this star is yet examined observationally.

In the current paper we presents the latest results of a long-term observational
programme aimed at the study of magnetic and chemical properties of three magnetic
stars: HD 178892, BD $+40^{\circ}175$ A and B. The paper is structured as follows.
In Sect. 2 we give the main information about the observational data and their
reduction. In the same section an observational log is placed as well. Section 3
contains the results of magnetic field measurements. In Sect. 4 we describe the
changes of the chemical composition of HD 178892 with the rotational phase and compare the
chemical compositions of both components of BD $+40^{\circ}175$ with each
other. Finally, in Sect. 5 we make some conclusions from our work and discuss the obtained results.

\section{Observations and data reduction}
For the purposes of our work we have used two devices: the Main Stellar Spectrograph~(MSS) and the Nasmyth Echelle Spectrometer~(NES), both installed in the Nasmyth-2 focus of the 6-m telescope of the Special Astrophysical Observatory of the Russian Academy of Sciences.

MSS, a long-slit spectrograph, is equipped with a circular polarization analyzer combined with an effective image slicer and rotatable quarter-wave retarder that is able to take fixed positions at 0 and 90$^{\circ}$~(Kudryavtsev et al. 2006). It is necessary to stress the fact that our instrumental setup is practically the same as in the earlier paper by Ryabchikova et al.~(2006) with the study of HD 178892, except for one change. From May 2010 we started to use a new detector based on the EEV CCD 42-90. Due to the larger size of the CCD ($4600{\times}2048$ pixels vs. $2048{\times}2048$ pixels for the old CCD 42-40) and equal pixel size we have extended the spectral range
for one exposure to a 
factor of two. Taking into account overlapping wavelength regions of our spectra we can state the compatibility of our measurements of the longitudinal stellar magnetic field presented hereafter 
for HD 178892 with those obtained earlier by other scientists with the MSS device of the Russian 6-m telescope.

The Nasmyth Echelle Spectrometer~(Panchuk et al. 2009) was used with a slit unit containing  an image slicer. The working spectral range was approximately the same as in the paper by Ryabchikova~(2006), the spectral resolution $R$ was about 40\,000. Such an instrumental setup also guarantees a high compatibility with older results. Details about  heliocentric Julian dates of the observations, spectral ranges, signal-to-noise ratios, mean resolving powers of the spectra, and names of the spectral devices are given in Table~\ref{table:1}.

\begin{table}
\caption{An observational log for the stars HD 178892 and BD\,$+40^{\circ}175$A and B.}
\label{table:1}
\begin{tabular}{c c c c c}\hline\noalign{\smallskip}
HJD & $\Delta\lambda$ (\AA) & S/N & $R$ & Device \\[1.5pt] \hline\noalign{\smallskip}
\multicolumn{5}{c}{HD 178892} \\
2 453 871.363 & 5576--7075 & 130 & 40 000 & NES \\
2 454 308.360 & 4385--5860 & 130 & 41 000 & NES \\
2 454 309.454 & 4385--5860 & 140 & 41 000 & NES \\
2 454 318.289 & 4385--5860 & 150 & 40 000 & NES \\
2 454 338.313 & 4385--4625 & 160 & 13 000 & MSS \\
2 454 395.170 & 4740--6168 & 150 & 39 000 & NES \\
2 454 396.167 & 4740--6168 & 140 & 39 000 & NES \\
2 454 397.174 & 4740--6168 & 140 & 39 000 & NES \\
2 454 488.632 & 4760--5000 & 150 & 13 000 & MSS \\
2 454 522.593 & 4765--5000 & 150 & 13 000 & MSS \\
2 454 610.394 & 4760--5002 & 140 & 12 000 & MSS \\
2 454 669.367 & 4758--4998 & 140 & 12 000 & MSS \\
2 454 702.252 & 4740--6170 & 120 & 40 000 & NES \\
2 454 955.424 & 4766--5005 & 160 & 13 000 & MSS \\
2 455 015.335 & 4765--5003 & 140 & 13 000 & MSS \\
2 455 460.236 & 4382--4938 & 150 & 15 000 & MSS \\
2 455 461.269 & 4382--4938 & 160 & 15 000 & MSS \\[5pt]
\multicolumn{5}{c}{BD $+40^{\circ}175$A} \\
2 454 396.451 & 4740--6168 &\enspace 90 & 39 000 & NES \\
2 454 702.387 & 4740--6168 & \enspace80 & 41 000 & NES \\
2 454 749.574 & 4740--6168 &\enspace 90 & 40 000 & NES \\
2 454 783.191 & 4215--4456 & 150 & 13 000 & MSS \\
2 455 784.473 & 4420--4975 & 120 & 14 000 & MSS \\
2 455 788.408 & 4420--4975 & 110 & 14 000 & MSS \\[5pt]
\multicolumn{5}{c}{BD $+40^{\circ}175$B} \\
2 454 396.514 & 4740--6168 &\enspace 75 & 39 000 & NES \\
2 454 700.561 & 5334--6788 &\enspace 55 & 41 000 & NES \\
2 454 702.434 & 4740--6168 &\enspace 70 & 41 000 & NES \\
2 454 783.302 & 4215--4456 & 130 & 13 000 & MSS \\
2 455 788.467 & 4420--4975 & 110 & 14 000 & MSS \\[1.5pt]
\hline
\end{tabular}
\end{table}

A common set of observational data collected with MSS consists of bias frames,
ThAr frames, spectra of a known magnetic standard star and zero-field standard
star(s), and spectra of target objects. Spectra of standard stars are necessary
for accounting the instrumental polarization. As a
`field'-standard star we usually used one or more magnetic stars with a well
measured $B_\mathrm{\rm e}$ curve. Reduction of the raw polarized spectra was done using
a set of programmes written by D.~Kudryavtsev for the {M}unich {I}mage
{D}ata {A}nalysis {S}ystem~(Kudryavtsev 2000; Kudryavtsev et al. 2006). 

The reduction includes the following next stages: computing of a master bias frame, bias correction, scattered light correction, creation of a mask with individual traces for slices,  extraction of one-dimen\-sional polarized spectra, wavelength calibration, and the correction of radial velocities. Actually, each individual stellar spectrum results from the combination of two exposures when the quarter-wave plate takes a fixed angle 0 and 90$^{\circ}$ relative to the beamsplitter. This allows to exposure circularly polarized spectra of opposite direction on the same pixels, and thus to eliminate effects related to the instrumental polarization of the spectrograph,  nonuniform response of CCD, etc.  Moreover, such an observational strategy allows to divide one long exposure in two shorter ones 
 in the case of faint stars, in order to reduce the impact from cosmic particles on the spectrograms.

The normalization of the spectra by their continuum level was  usually done with the task {\sc continuum} from the {I}mage {R}edu\-ction and {A}nalysis {F}acility\footnote{IRAF is distributed by the National Optical Astronomy Observatories, which are operated by the Association of Universities for Research in Astronomy, Inc., under cooperative agreement with the National Science Foundation.}. However, this last procedure is not obligatory for the application of Zeeman shift measurements. 

The reduction of \'echelle spectra was carried out with a set of IDL routines called
REDUCE~(Piskunov \& Valenti 2002). The raw data were processed in a standard way, including subtraction of average bias, flat field normalization and subtraction of stray light. As in the case of polarized long slit spectra, all \'echelle spectra were manually normalized by the continuum level with the task {\sc continuum}. For each spectrum the normalization was performed order-by-order with a smoothed g of blaze function of the spectrograph. Thus we minimize the impact of the edge effects caused by lower SNR and camera's vignetting on the final spectrograms.

\section{Magnetic fields of the target stars}
\subsection*{HD 178892}
The decision to continue the spectral and spectropolarimetric observations of HD 178892 was taken immediately after the publication of Ryabchikova et al.~(2006). These new observations were aimed to  study in detail the variation of the magnetic properties with rotation.

\begin{figure}
\vskip1mm
\hskip 4mm
\includegraphics[width=72mm]{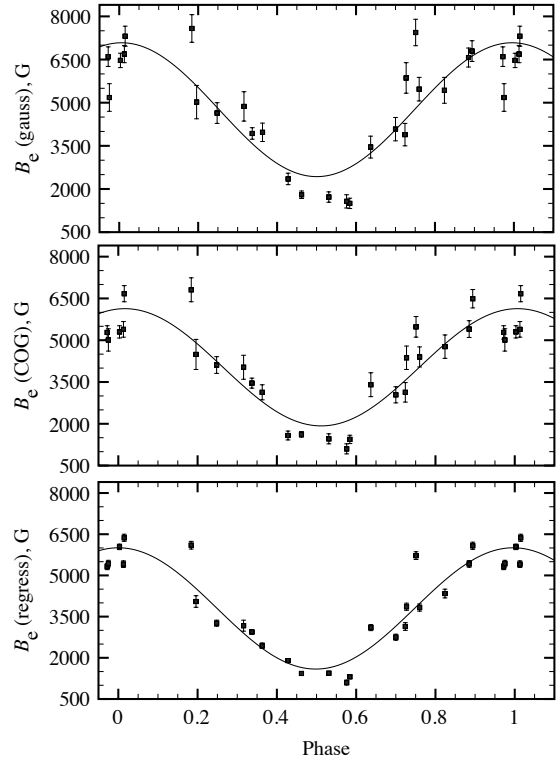}
\vskip 2mm
\caption{Variation of the longitudinal magnetic field of HD 178892, phased with a rotational period of 8.2549 days as measured using three different methods.}
\label{figure:1}
\end{figure}

During the last 5 years we have obtained in total 8 new \'echelle  and 9 long-slit circularly polarized spectra. These new spectrograms are distributed rather well along the known $B_\mathrm{\rm e}$ curve. However, discrepancies between our new measurements and the curve  computed with the assumption of a period of 8.2478 days forced us to re-determine the rotational period of this star. The new value of 8.2549 days were determined using a phase dispersion minimization algorithm  implemented in the task {\sc pdm} from the IRAF package {\sc astutil}. It is interesting that the close period  of 8.25 days also was mentioned in the paper by Ryabchikova et al.~(2006). Figure~\ref{figure:1} shows the phased variation of the longitudinal magnetic field of the star based on  the ephemeris $\mathrm{HJD} = 2\,452\,708.562+8.2549{\times}E$.

This figure represents the longitudinal magnetic field of HD\,178892 as it was measured by three different methods. For this figure all spectra registered before the end of 2006 were re-measured as well. The upper and middle sets of points were obtained by a method where the value of $B_\mathrm{e}$ is computed as follows: 
\begin{displaymath}
B_\mathrm{e}=\frac{\Delta\lambda}{9.34{\times}10^{-13}\,\lambda_{0}^{2}\,g_\mathrm{eff}}
\quad\mathrm{[G]},
\end{displaymath}
where $\Delta\lambda$ is the difference between the $\sigma$ components of a line in circularly polarized spectrograms. In the first case~(upper plot) the position of individual lines in each polarization was derived from a Gaussian fit. This method is common for most of the measurements taken by our research group.

The middle plot contains measurements received in a similar way as before, but the center of individual lines was determined  using the center-of-gravity method, according to the
 relation
\begin{displaymath}
\lambda_\mathrm{cog}=\frac{\int \lambda(I_\mathrm{cont}-I)\,\mathrm{d}\lambda}{\int (I_\mathrm{cont}-I)\,\mathrm{d}\lambda}.
\end{displaymath}
Thus, $\Delta\lambda=\lambda_\mathrm{cog}(\rm RCP)-\lambda_\mathrm{cog}(LCP)$.

The third method we used in this work is the linear regression method which was described in a paper by Bagnulo et al.~(2002).

The comparison of the three different $B_\mathrm{e}$ curves shows a significant scattering of individual points from the sinusoidal fit. The best fitted set of points is on the bottom panel, but the errors from this method seem to be underestimated. Attention is drawn to the fact that the maximum values of the longitudinal field in middle and bottom panel are nearly the same and significantly less than that in the top panel. Also there are two spikes corresponding to the phase of about 0.2 and 0.75. The origin of these spikes is rather unknown. A careful study of these two spectrograms has shown the absence of any essential distinctions to spectra taken in similar phases. In order to clarify the nature of these outliers it is necessary to obtain new spectrograms at nearby rotational phases.

The long standing Zeeman measurements that took pla\-ce in our observatory confirm the validity of the simplified determination of line centers using Gaussian fits. But this method meets a limitations when we deal with strongly magnetic stars. The Zeeman effect in stellar spectra distort the shape of individual lines and can lead to partial line splitting. Thus,  the observed overestimation of the maximum values of the magnetic field in the upper panel of Fig.~\ref{figure:1} may be due to the influence of the Zeeman effect on the line fitting algorithm. Under such circumstances the determination of  the line center is not straightforward. We believe that the center-of-gravity method is more stable even in the case of strong fields. That is why we consider the magnetic curve in the middle panel of Fig.~\ref{figure:1} as more reliable. It is interesting that, while the weak-field regime in the case of HD 178892 is obviously broken, the linear regression method still works well.

\begin{figure}
\hskip -2mm
\includegraphics[width=82mm]{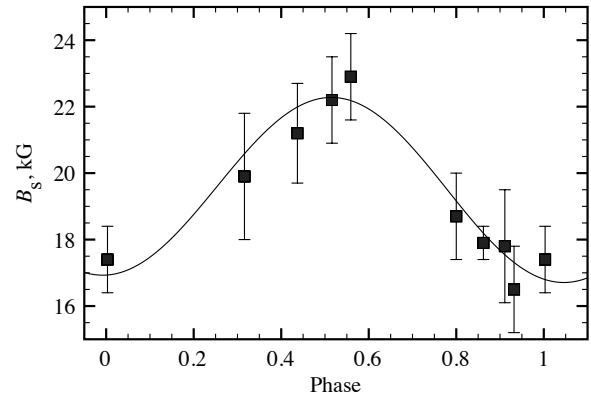}
\vskip-2mm
\caption{Variation of the mean modulus of the surface magnetic field of HD 178892 phased  with the rotational period of 8.2549 days. The scale of the abscissa is the same as in Fig.~\ref{figure:1}.}
\label{figure:2}
\end{figure}

Any detailed analysis of magnetic fields would be incomplete without involving  measurements of the average surface magnetic field $B_\mathrm{s}$. Ryabchikova et al.~(2006) in their work have showed that the surface magnetic field of HD 178892 is large but they had no information about its variability. We have filled this gap using eight new \'echelle spectra. Measurements were performed with a the set of spectral lines with distinctive  resolved Zeeman patterns. All measurements of the surface magnetic field including points from the paper of 2006 are displayed in Fig.~\ref{figure:2}.

Comparison of Fig.\,\ref{figure:2} with any part of Fig.\,\ref{figure:1} shows that the longitudinal and surface magnetic field vary in antiphase. Moreover, $B_\mathrm{s}$ reaches its maximum by about 0.1 of the period early than $B_\mathrm{e}$~ its minimum. These leading phase shift and anti-phase changes of the components of a magnetic field can indicate the complex structure of a stellar field. An attempt of measuring split line components belonging to different species gives different results. The observed differences could be a result of an inhomogeneous distribution of chemical elements over the stellar surface. However, since  many of lines are not always present in each spectrogram we do not present here the results referring to individual elements.

\subsection*{BD $\mathbf{+40^{\circ}175}$AB}

The magnetic nature of the main component of this binary star was established for the first time by Babel \& North~(1997). Though the observations of this star are very complicated~($m^\mathrm{A}_{V}=9.5$ mag, $m^\mathrm{B}_{V}=9.9$ mag, $\rho\approx 3\farcs7$), the authors were able to measure a mean modulus of the surface magnetic field of the main component
by means of a set of \'echelle spectrograms.
According to Babel \& North, the surface magnetic field of BD $+40^{\circ}175$A varies from 
about 12 kG up to more than 22 kG. However, the \'echelle spectrograms used in their work are of low signal-to-noise ratios. In order to improve the quality of the results, the authors have applied a correlational analysis aimed at the derivation of a mean line profile.

For a long time after the paper of Babel \& North~(1997) no studies of the star with high-resolution spectroscopy has been carried out. But Elkin~(1999), who observed both components of this binary using the 6-m Russian telescope, has found that the second component is a magnetic object too. In spite of the previous work, Elkin used the circular polarized analyzer and medium spectral resolution. This allowed him to measure the longitudinal magnetic field of both components with a reasonable accuracy,
based on five measurements of $B_\mathrm{e}$ for each component. These measurements revealed fast variations of $B_\mathrm{e}$, from $-$2.2 kG to $-$3.4 kG, and from 0.7 to 2.6 kG for the components A and B, respectively. That is a sign that the longitudinal field is opposite! Therefore, we have a unique binary star where both components are magnetic and where the main component possesses an exceptionally strong surface magnetic field.

We started our study of this binary system in 2007.  We directed our main efforts to the analysis of \'echelle spectrograms, but  Zeeman measurements were carried out as well. As a result we got three additional $B_\mathrm{e}$ points for the component A, and two points~for the component B.
The situation with the measurements of a surface magnetic field was similar. The longitudinal magnetic field was measured using the method of Gaussian fitting, as described in the previous subsection.

Since the double system is only occasionally observed, the rotational periods of its components are not known. We collected together all available data on the magnetic field of BD $+40^{\circ}175$ AB in Table~\ref{table:2}. 

\begin{table}
\caption{Magnetic field measurements for both components of BD $+40^{\circ}175$.}
\label{table:2}
\begin{tabular}{c c c c}
\hline\noalign{\smallskip}
HJD &    $B_\mathrm{e}\pm\sigma$ (G) & $B_\mathrm{s}\pm\sigma$ (kG) & Ref. \\[1.5pt]
\hline\noalign{\smallskip}
\multicolumn{4}{c}{BD $+40^{\circ}175$A} \\
2 450 118.248  &  & $12.8\phantom{\pm}1.5$ & 1 \\
2 450 118.318  &  & $12.7\phantom{\pm}1.5$ & 1 \\
2 450 119.251  &  & $16.1\phantom{\pm}1.5$ & 1 \\
2 450 122.252  &  & $15.7\phantom{\pm}1.5$ & 1 \\
2 450 349.459  &  & $22.3\phantom{\pm}2.5$ & 1 \\
2 450 709.581  & $-2020\phantom{\pm}215$ & & 2 \\
2 450 710.322  & $-3330\phantom{\pm}170$ & & 2 \\
2 450 774.367  & $-2750\phantom{\pm}205$ & & 2 \\
2 450 829.262  & $-3400\phantom{\pm}180$ & & 2 \\
2 451 064.316  & $-2840\phantom{\pm}260$ & & 2 \\
2 454 396.451  &   & $13.8\phantom{\pm}1.5$ & 3 \\
2 454 708.387  &   & $14.4\phantom{\pm}1.0$ & 3 \\
2 454 749.574  &   & $14.3\phantom{\pm}0.8$ & 3 \\
2 454 783.191  & $-2430\phantom{\pm}130$ & & 3 \\
2 455 784.473  & $-1470\phantom{\pm}145$ & & 3 \\
2 455 788.408  & $-3310\phantom{\pm}\enspace90$  & & 3 \\[5pt]
\multicolumn{4}{c}{BD $+40^{\circ}175$B} \\
2 450 709.581  &\enspace \enspace$780\phantom{\pm}215$ &  & 1  \\
2 450 710.303  &\enspace $1050\phantom{\pm}\enspace80$ &  & 1  \\
2 450 774.389  &\enspace $1540\phantom{\pm}205$ &  & 1 \\
2 450 829.285  &\enspace $1300\phantom{\pm}120$ &  & 1 \\
2 451 064.340  &\enspace $2660\phantom{\pm}190$ &  & 1 \\
2 454 396.514  &   & $11.7\phantom{\pm}1.0$ & 3 \\
2 454 700.561  &   & $11.9\phantom{\pm}1.8$ & 3 \\
2 454 702.434  &   & $\enspace9.6\phantom{\pm}0.4$  & 3 \\
2 454 783.302  &\enspace $1245\phantom{\pm}160$ & & 3 \\
2 455 788.467  &\enspace $\enspace960\phantom{\pm}\enspace70$ &  & 3 \\[1.5pt]
\hline
\end{tabular}
\\[3pt]
Ref.: (1) Babel \& North (1997); (2) Elkin (1999); (3) this work.
\end{table}

\begin{figure}
\centering\includegraphics[width=82mm]{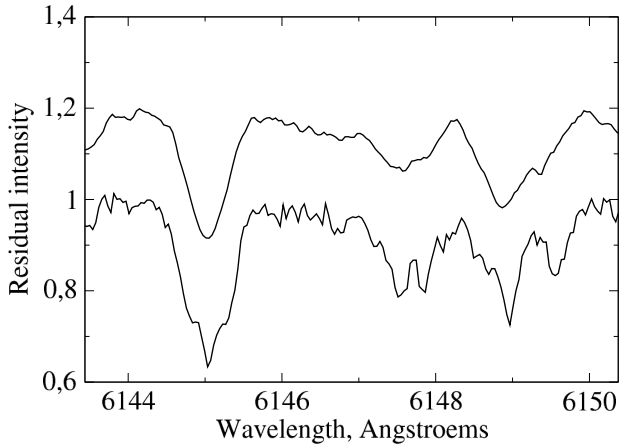}
\caption{Spectra of the components A (\emph{bottom}) and B (\emph{top}) in the region of Nd\,{\sc iii} 6145 \AA\ and Fe\,{\sc ii}6149 \AA. The spectrum of the secondary component is artificially shifted upward by 0.2.}
\label{figure:3}
\end{figure}

The data from Table~\ref{table:2} show that the magnetic field of the secondary component is slightly lower than that of component A. Due to the slow rotation of component A~($v\sin i=9.7\pm2.6$ km\,s$^{-1}$), the spectrum is rich in lines with resolved Zeeman components. The secondary star rotates faster, its projected rotational velocity $v\sin i$ is equal to $17.4\pm2.9$ km\,s$^{-1}$,  but some  resolved lines are still detectable~(Fig. \ref{figure:3}).

It can be seen from the Table~\ref{table:2} that our measurements are in a good agreement with those  from Babel \& North (1997). However, none of our $B_\mathrm{s}$ values exceed 20 kG for either components. This fact may be explained by a stochastic coincidence of the phase when the magnetic field were non-maximal during the time of observations. Another explanation is a poor quality of the spectral material used by Babel \& North. The value of 22.3 kG with a rather large error of 2.5 kG testifies in favor of 
our assumption, also because the other results from Babel \& North  agree well with our data.

Our measurements of the longitudinal and surface magnetic fields have continued the study of this unique double star started by Babel \& North~(1997) and Elkin~(1999). However, there remain many
uncertainties concerning the properties of this binary. In order to finally clarify the magnetic properties of 
BD\,$+40^{\circ}175$ it is needed to carry out more observations with the largest telescopes.

\begin{table*}
\caption{Element abundance of the atmospheres of  HD 178892, BD $+40^{\circ}175$A and B. For comparison, the last column contains the solar abundances from Asplund et al.~(2005).}
\label{table:3}
\tabcolsep=10pt
\begin{tabular}{l c c @{\hspace{2.0cm}} c c  r}
\hline\noalign{\smallskip}
      & \multicolumn{5}{ c }{$\log (N/N_\mathrm{tot})\ (\sigma)$} \\[1.5pt]
\cline{2-6}\noalign{\smallskip}
 Ion & \multicolumn{2}{ c }{HD 178892} & \multicolumn{2}{c }{BD $+40^{\circ}175$} & Sun~~ \\[1.5pt]
\cline{2-5}\noalign{\smallskip}
      & \multicolumn{1}{ c }{$B_\mathrm{s}=16.5$ kG} & \multicolumn{1}{l }{$B_\mathrm{s}=22.9$ kG} & \multicolumn{1}{c }{Primary Comp.} & \multicolumn{1}{c }{Second. Comp.} &    \\[1.5pt]
\hline\noalign{\smallskip}
C {\sc i}    & $-3.75\ (0.16)$ & $-3.83\ (0.07)$ &   &  &  $-3.45$ \\
Si {\sc i}   & $-3.35\ (0.15)$ & $-3.67\ (0.18)$ &   &  &  $-4.50$ \\
Si {\sc ii}  & $-3.29\ (0.10)$ & $-3.37\ (0.10)$ & $-3.80\ (--)$ & $-3.45\ (0.25)$ & $-4.50$ \\
Ti {\sc i}   & $-6.66\ (--)$      &    &  &    &   $-7.02$ \\
Ti {\sc ii}  & $-6.70\ (0.25)$  & $-7.35\ (--)$ &   &   &  $-7.02$ \\
Cr {\sc i}  & $-5.25\ (0.18)$  & $-5.82\ (--)$ &   &   &  $-6.37$ \\
Cr {\sc ii} & $-5.44\ (0.38)$  & $-5.69\ (0.15)$ & $-5.48\ (0.10)$ & $-5.20\ (--)$ & $-6.37$ \\
Fe {\sc i}   & $-5.29\ (0.13)$ & $-5.43\ (0.29)$ & $-4.70\ (0.26)$ & $-4.46\ (0.08)$ & $-4.59$ \\
Fe {\sc ii}  & $-5.24\ (0.25)$ & $-5.44\ (--)$ & $-4.85\ (0.05)$ & $-4.30\ (0.10)$ & $-4.59$ \\
Y {\sc ii}   & $-9.25\ (0.18)$ & $-9.42\ (0.24)$  &   &   &  $-9.80$ \\
La {\sc ii}  & $-8.86\ (0.26)$ &   &    &   &   $-10.87$ \\
Ce {\sc ii}  & $-8.84\ (0.13)$ & $-8.62\ (0.29)$ &   &   &  $-10.46$ \\
Pr {\sc ii}  & $-8.95\ (0.24)$ & $-9.03\ (0.27)$  & $-8.55\ (0.35)$ & $-8.00\ (--)$ & $-11.33$ \\
Pr {\sc iii} & $-7.87\ (0.20)$ & $-7.91\ (0.38)$  &   &  $-7.30\ (--)$  & $-11.33$  \\
Nd {\sc ii} & $-8.45\ (0.21)$ & $-8.51\ (0.29)$  & $-8.44\ (0.29)$ & $-8.10\ (0.10)$ & $-10.54$ \\
Nd {\sc iii} & $-7.05\ (0.10)$ & $-7.34\ (0.22)$ & $-7.25\ (0.25)$ & $-6.80\ (--)$ & $-10.54$ \\
Sm {\sc ii} & $-8.56\ (0.21)$ & $-8.59\ (0.22)$ &    &    &   $-11.03$ \\
Eu {\sc ii}  &   & $-8.36\ (0.32)$ &   &   &  $-11.53$  \\
Tb {\sc ii}  & $-9.90\ (--)$ &    &    &    &   $-12.14$  \\
Tb {\sc iii} & $-7.61\ (0.07)$ & $-7.76\ (0.42)$ &   & $-7.60\ (--)$ & $-12.14$ \\[1.5pt]
\hline
\end{tabular}
\end{table*}

\section{Elemental abundances of the target stars}

The chemical abundances of the atmospheres of HD 178892 and BD $+40^{\circ}175$ have been estimated already by other researchers. However, in the case of HD 178892 we have decided to determine the abundances of several elements using the spectra that correspond to minimum and maximum of the surface field. We wanted to study how the magnetic field can effect individual abundances. Regarding the binary star BD $+40^{\circ}175$, a detailed analysis of high resolution spectrograms has never been performed yet.

For  HD 178892  we have used two spectrograms which correspond to the phases 0.55 and 0.93 where the magnetic field reach its extrema. Individual abundances of both components of the double star BD $+40^{\circ}175$ were extracted from the spectra obtained at HJD = 2\,454\,396.

Atmospheric parameters of HD 178892 were taken from the paper of Ryabchikova et al.~(2006). According to this paper, the temperature $T_\mathrm{eff}$ is 7700 K and the surface gravity $\log g$ = 4.0. Since for BD $+40^{\circ}175$ there do not exist any estimates of the atmospheric parameters we tried to determine them ourself. This binary star was, due to its faintness, practically not observed before. 
We have determined the atmospheric parameters from the hydrogen lines profiles. Comparing the observed line profiles of H$\beta$ and H$\gamma$ with  theoretical predictions we have found the best agreement when $T_\mathrm{eff}=7700$ K, $\log g=4.0$, and if the metallicity [M/H] is close to $+0.5$. These parameters describe the observed hydrogen lines of the component A fairly well, but the secondary component appears to be a bit cooler. Thus, assuming an error of the effective temperature of about 350 K we  expect that the real temperature of the secondary component could be within this range.

All the analyses were done in the LTE approach. In order to take into account the effects of stellar magnetic field we used the {\sc SynthMag} code~(Kochukhov 2007). In all particular cases the magnetic field was taken equal to the value of the surface field, as measured from the observed spectra.
Final abundances for all studied stars are gathered in Table~\ref{table:3}.

\section{Discussion}
In the current paper we presents the latest results for three strongly magnetic stars. the first object is the famous Ap star HD\,178892. A previous study carried out by Ryabchikova et al. (2006) has revealed the presence of an extraordinary strong magnetic field on the stellar surface. However, this result (${B_\mathrm{s}=17.5}$ kG) was based on a small number of individual measurements corresponding to nearby phases. In the current work we have extended the set of measurements by adding new eight data points. Due to these new data we can state that the magnetic field of HD 178892 is more complex and stronger than previously thought.

We have measured all available circularly polarized spectrograms of the star using
the modified positional method. Applying the center-of-gravity method to
measure the positions of individual lines, we found that the maximum value of the
stellar longitudinal field is about one kilogauss less than given in the paper by
Ryabchikova et al. (2006). Also we have found that the maximum of the surface field is
reached when the longitudinal field is about at its minimum. Using a dipolar model of the
magnetic field we were able to estimate the main parameters of the dipole. Assuming
the stellar radius of HD 178892 to be equal to $1.75\,$M$_{\odot}$~(Harmanec 1988), and
taking a rotational period of 8.2549 days, the rotational velocity on the stellar equator should
be about 10 km\,s$^{-1}$. According to Ryabchikova et al., the projected rotational
velocity of the star is about 9~km\,s$^{-1}$. Therefore, the angle $i$ between
rotational axis of the star and the direction towards an observer should be about
57\fdg6. 

From the  $B_\mathrm{s}$ curve presented in Fig.~\ref{figure:2} we can estimate the maximum and minimum values of the surface magnetic field: 17 and 22.5 kG, respectively. At the same time the longitudinal magnetic field varies from 2 to 6 kG (middle panel of Fig.~\ref{figure:1}). Thus, assuming the angle between rotational and dipole axes to be $21^{\circ}$, the strength of the magnetic field on the pole of the dipole, $B_\mathrm{p}$, is equal to 28 kG. In this sense HD 178892 is very similar to another cool Ap star with a strong magnetic field -- HD 154708~(Hubrig et al. 2005, 2009). 

This similarity  increase when we look also on the  chemical compositions. We have carried out a detailed study of the chemical abundances in the atmosphere of HD 178892. As a result, we obtained two samples of data corresponding to the moments where the stellar magnetic field achieves its extrema. Generally, our results are close to those from the paper by Ryabchikova et al. (2006). The star has an abundance pattern typical for roAp stars with a significant Pr-Nd anomaly. At the moment when $B_\mathrm{s}$ is near its maximum, individual abundances are not changed. This means that HD 178892 may  have a chemically homogeneous distribution of the main chemical elements over the surface. The strong spectral variability noted by Ryabchikova et al. may be caused by magnetic intensification.

Constructing a model of the magnetic field of HD 178892 we do not take into account the fact of the antiphased variation of the field. The above-mentioned calculations of magnetic models were based on the assumption a simple central dipole configuration. In order to describe the real behavior of both magnetic curves, ($B_\mathrm{e}$ and $B_\mathrm{s}$), we should consider more complex structures of the field, including a multipole expansion.

The  outliers in the individual $B_\mathrm{e}$ measurements found near phases 0.2 and 0.75 could be connected with geometrical effects; approximately at these phases we have the maximum of the cross-over effect.

Concerning the comparison of HD 178892 and HD 154708 as two most outstanding cool magnetic stars, it would be interesting to look more detailed at the chemical composition of the second star. We found only a brief paper by Nesvacil et al.~(2008) where individual abundances are presented as a diagram. Also, in spite of the study  by Hubrig et al.~(2009) about HD 154708's field geometry, all their conclusions about the surface magnetic field were based on a model, while real measurements of the modulus of the surface field in different phases are absent.

Considering the binary star BD $+40^{\circ}175$, our paper is a first attempt to carry out a study of this unique stellar system using high-resolution spectrograms. We have extended the total sample of $B_\mathrm{e}$ and $B_\mathrm{s}$ measurements for both components. For the secondary star direct measurements of $B_\mathrm{s}$ were obtained for the first time. Unfortunately, we cannot find any references to the rotational period of these stars. These data are essential for modeling the magnetic field geometry. 

Our estimation of the chemical composition of the stellar atmospheres has shown a strong anomaly in Pr, Nd, and Tb that is typical for rapidly oscillating stars. Both components have a temperature that makes  BD $+40^{\circ}175$A and B  good candidates for roAp stars. Taking into account the strong magnetic field exceeding 10 kG, we have a unique chance to check how common strong magnetic fields are among roAp stars.

Despite  the wide angular separation of the components and at the same time their faintness, we believe that this system is a gravitationally coupled one. Thus the study of the magnetic field, chemical composition and other stellar parameters of systems similar to BD\,$+40^{\circ}175$  could give us an essential information about the origin and evolution of magnetic CP stars.

\acknowledgements
Authors are thankful to Dr. D. Kudryavtsev who assisted in obtaining the observations, and to Drs. I. Romanyuk and T. Ryabchikova for useful discussions of the results. The current work was partially sponsored by the Russian Foundation for Basic Research (grant No.\, 09-02-00002-a. The study of HD 178892 was carried out with financial support from the Ministry of Education and Science of Russian Federation within the Federal programme ``Scientific and educational cadre of innovating Russia 2009--2013'' (grant No.\,P1194).

\end{document}